\documentclass[preprint,prd,nofootinbib,tightenlines,groupedaddress,amsmath,amssymb]{revtex4}
\usepackage{bm}
\usepackage{color}
\usepackage{graphicx}
\allowdisplaybreaks
\usepackage{multirow}
\begin{document}
\title{Model of Neutrino Mass Matrix With $\delta = -\pi/2$ and $\theta_{23} = \pi/4$}
\author{Xiao-Gang He$^{1,2,3}$\footnote{hexg@phys.ntu.edu.tw}}
\affiliation{${}^{1}$INPAC, SKLPPC and Department of Physics,
Shanghai Jiao Tong University, Shanghai, China}
\affiliation{${}^{2}$CTS, CASTS and
Department of Physics, National Taiwan University, Taipei, Taiwan\\}
\affiliation{${}^{3}$National Center for
Theoretical Sciences and Physics Department of National Tsing Hua University, Hsinchu, Taiwan}
\date{\today $\vphantom{\bigg|_{\bigg|}^|}$}
\date{\today}
\vskip 1cm
\begin{abstract}
Experimental data have provided stringent constraints on neutrino mixing parameters. In the standard parameterization the mixing angle $\theta_{23}$ is close to $\pi/4$. There are also evidences show that the CP violating phase is close to $-\pi/2$. We study neutrino mass matrix reconstructed using this information and find several interesting properties. We show that a theoretical model based on the $A_4$ symmetry naturally predicts $\delta = -\pi/2$ and $\theta_{23} = \pi/4$ when the Yukawa couplings and scalar vacuum expectation values are real reaching a $\mu-\tau$ exchange and CP conjugate symmetry limit. In this case CP violation solely comes from the complex group theoretical Clebsh-Gordan coefficients. The model also predicts $|V_{e2}|=1/\sqrt{3}$ consistent with data. With complex Yukawa couplings the values for $\delta$ and $\theta_{23}$ can be significantly deviate away from the symmetry values $-\pi/2$ and $\pi/4$, respectively. But $|V_{e2}|= 1/\sqrt{3}$ is not altered. This matrix is an excellent lowest order approximation for theoretical model buildings of neutrino mass matrix.
\end{abstract}
\pacs{PACS numbers: }
\maketitle
Tremendous experimental progresses have been made in obtaining information about the neutrino mixing parameters. The mixing angles in the Pontecorvo-Maki-Nagakawa-Sakata $V_{PMNS}$ matrix\cite{pmns} are not always small\cite{pdg,valle,fogli-schwetz}. In the standard parameterization\cite{pdg, ckm} for three neutrino mixing commonly used\cite{valle,fogli-schwetz}, the mixing angle $\theta_{23}$ is close to $\pi/4$, $\theta_{12}$ is large, $\theta_{13}$ is relatively small but away from zero, and also $s_{12} c_{13}$ is close to $1/\sqrt{3}$. Since the mixing angle $\theta_{13}$ is non-zero, the famous tri-bimaximal mixing\cite{tri-bi} is ruled out. There are now evidences show that the CP violating phase $\delta$ is close to $-\pi/2$. This also implies that the tri-bimaximal mixing is in trouble since it predicts $\delta =0$. The phase $\delta$ is sometimes referred as Dirac phase which shows up in neutrino oscillations. If neutrinos are Majorana particles, there are also new CP violating Majorana phases $\alpha_i$. There are many discussions about implications for data available emphasizing the particular values for $|\delta|=\pi/2$ and $\theta_{23} = \pi/4$\cite{ma-a4,grimus,xing}. One of the commonly mentioned property for this type of mixing is the so called maximal CP violation because $|\delta|$ is $\pi/2$. This is, strictly speaking, an incorrect statement because that the value of the Dirac phase is parametrization dependent. For example, even the absolute value of the Dirac phase is $\pi/2$ in the standard parametrization, in the original Kobayashi-Maskawa parametrization for quarks\cite{km} it is not $\pi/2$ anymore. However, the special values for some of the mixing angles and the Dirac phase can still provide important information about neutrino mass matrix and can guide theoretical model buildings to search for the underlying theory. 

To this end, let us reconstruct the neutrino mass matrix assuming that neutrinos are Majorana particles with $\delta=-\pi/2$ and $\theta_{23} = \pi/4$. In the basis where the charged lepton mass matrix is already diaganolized, the neutrino mass matrix defined by the term giving neutrino masses in the Lagrangian
$(1/2)\bar \nu_L m_\nu \nu^c_L$, has the following form
\begin{eqnarray}
m_\nu = V_{PMNS} \hat m_\nu V^T_{PMNS}\;,
\end{eqnarray}
where $\hat m_\nu = diag(m_1, m_2, m_3)$ with $m_i = |m_i|exp(i\alpha_i)$. Here we have put Majorana phase information in the neutrino masses. The standard form for $V_{PMNS}$ is given by
\begin{eqnarray}
V_{PMNS} = \left(\!\begin{array}{ccc}
c_{12\,}^{}c_{13}^{} & s_{12\,}^{}c_{13}^{} & s_{13}^{}\,e^{-i\delta}
\vspace{1pt} \\
-s_{12\,}^{}c_{23}^{}-c_{12\,}^{}s_{23\,}^{}s_{13}^{}\,e^{i\delta} & ~~
c_{12\,}^{}c_{23}^{}-s_{12\,}^{}s_{23\,}^{}s_{13}^{}\,e^{i\delta} ~~ & s_{23\,}^{}c_{13}^{}
\vspace{1pt} \\
s_{12\,}^{}s_{23}^{}-c_{12\,}^{}c_{23\,}^{}s_{13}^{}\,e^{i\delta} &
-c_{12\,}^{}s_{23}^{}-s_{12\,}^{}c_{23\,}^{}s_{13}^{}\,e^{i\delta} & c_{23\,}^{}c_{13}^{}
\end{array}\right) ,
\label{pmns}
\end{eqnarray}
where $c_{ij}$ and $s_{ij}$ are $\cos\theta_{ij}$ and $\sin\theta_{ij}$, respectively. 
With $\delta = -\pi/2$ and $\theta_{23}=\pi/4$, $m_\nu$ has the following form\cite{ma-a4,xing}
\begin{eqnarray}
m_\nu = \left (\begin{array}{ccc}
a& c+i\beta&-(c - i\beta)\\
c+i\beta&d+i\gamma&b\\
-(c-i\beta)&b&d-i\gamma
\end{array}
\right )\;,
\label{matrix}
\end{eqnarray}
where
\begin{eqnarray}
&&a=m_1c^2_{12} c^2_{13}+m_2s^2_{12} c^2_{13}-m_3 s^2_{13}\;,\;\;\;\;\;\;\;\;\;b =-\frac{1}{2} \left(m_1(s^2_{12} + c^2_{12} s^2_{13}) +m_2(c^2_{12}+ s^2_{12}s^2_{13})-m_3 c^2_{13}\right)\;,\nonumber\\
&&c=- \frac{1}{\sqrt{2}} (m_1-m_2) s_{12}c_{12} c_{13}\;,\;\;\;\;\;\;\;\;\;\;\;\;\;\;\;
d=\frac{1}{2} \left(m_1 (s^2_{12} - c^2_{12} s^2_{13})
+m_2 (c^2_{12} - s^2_{12} s^2_{13})+m_3 c^2_{13}\right )\;,\nonumber\\
&&\beta = \frac{1}{\sqrt{2}} s_{13} c_{13}\left(m_1 c_{12}^2+m_2 s^2_{12}+m_3\right)\;,\;\;\gamma = - (m_1-m_2) s_{12} c_{12} s_{13} \;.
\label{eeq}
\end{eqnarray}
Note that in the most general case, because non-zero Majorana phases, the parameters $a$, $b$, $c$, $d$, $\beta$ and $\gamma$ are all complex. 

The above matrix has a high level regularity pattern implying some underlying symmetry may be at work to produce it. Searching an underlying theory guided by symmetry principle may achieve this. Before doing this, however, it is worthwhile to understand more about the mass matrix in eq.(\ref{matrix}). An immediate question one may ask is that if, in general, the neutrino mass matrix in eq.(\ref{matrix}) always predicts $\delta = -\pi/2$ and $\theta_{23} = \pi/4$. The answer is negative. If $\delta = \pi/2$ and $\theta_{23} = \pi/4$, the neutrino mass matrix is given in a similar form as that in eq.(\ref{matrix}), but $\beta$ and $\gamma$ need to be multiplied by a ``-'' sign. Therefore without further information given, a general mass matrix in the form given by eq.(\ref{matrix}) can give $\delta = \pm \pi/2$ and $\theta_{23}=\pi/4$. Whether they predict $+\pi/2$ or $-\pi/2$, additional information need to be provided. 
Moreover, If neutrinos have Majorana phases, the general form does not imply that
$\delta$ and $\theta_{23}$ must take $\pm \pi/2$ and $\pi/4$, respectively, neither. This can be understood by studying the following quantity
\begin{eqnarray}
m_\nu m_\nu^\dagger = V_{PMNS} \hat m_\nu \hat m^\dagger_\nu V^\dagger_{PMNS}\;.
\end{eqnarray}
The general form for neutrino mass in eq.(\ref{matrix}) will give
the ``12'' and ``13'' entries $A_{12, 13}$ of $m_\nu m^\dagger_\nu$ as
\begin{eqnarray}
A_{12} +A_{13} &&= - i 2 (a \beta^* + c \gamma^* - \beta d^* -\beta b^*)\nonumber\\
&&=-(|m_1|^2-|m_2|^2) s_{12}c_{12}c_{13} ( c_{23}-s_{23} )\nonumber\\
&&\;\;\;\;- (|m_1|^2 c_{12}^2 + |m_2|^2s_{12}^2 - |m_3|^2) s_{13} c_{13} (c_{23}+s_{23} )e^{-i\delta}\;,\nonumber\\
A_{12} - A_{13} &&= 2 (a c^* + c d^*- c b^* + \beta \gamma^*)\nonumber\\
&&=-(|m_1|^2-|m_2|^2) s_{12}c_{12}c_{13} ( c_{23}+s_{23} )\nonumber\\
&&\;\;\;\;+ (|m_1|^2 c_{12}^2 + |m_2|^2s_{12}^2 - |m_3|^2) s_{13} c_{13} (c_{23}-s_{23} )e^{-i\delta}\;.
\label{general}
\end{eqnarray}

If the parameters in the set $ P$: \{$a$, $b$, $c$, $\beta$, $\gamma$\}, are complex, the above equations can find solutions for other values of $\theta_{23}$ and $\delta$. Therefore the general neutrino mass matrix form does not imply that $\delta$ and $\theta_{23}$ must be 
$\pm \pi/2$ and $\pi/4$. 
If, however, the parameters in the set $P$ are all real, as long as $\sin\delta \neq 0$, one must have $s_{23} = c_{23}$ and $\delta= \pm\pi/2$ as can be seen from the above two equations. From eq.(\ref{eeq}) and eq.(\ref{general}), one also finds that all eigen-masses $m_i$ are real (the Majorana phases are zero or $\pi$). In this case the neutrino mass matrix can be rewritten as 
\begin{eqnarray}
m_\nu = \left (\begin{array}{ccc}
A& C&-C^*\\
C&D^*&B\\
-C^*&B&D
\end{array}
\right )\;,
\end{eqnarray}
with $A = a$, $B=b$, $C = c+i\beta$, and $D=d-i\gamma$. 
The most general $m_\nu$ can be written as\cite{ma-a4}
\begin{eqnarray}
m_\nu &=& \left (\begin{array}{ccc}
e^{ip_1}&0&0\\
0&e^{ip_2}&0\\
0&0&e^{ip_3}
\end{array}\right )
\left (\begin{array}{ccc}
A& C&-C^*\\
C&D^*&B\\
-C^*&B&D
\end{array}
\right )\left (\begin{array}{ccc}
e^{ip_1}&0&0\\
0&e^{ip_2}&0\\
0&0&e^{ip_3}
\end{array}\right )\;,
%
\end{eqnarray}
where the phases $p_i$ are arbitrary. 

All neutrino mass matrices which can be written in the above form, will predict
$\delta = \pm\pi/2$, $\theta_{23} = \pi/4$ and all the eigen-masses are real. One can choose some particular values for $p_i$ to obtain forms of $m_\nu$ for convenience of analysis. For example the ``-'' sign for the ``13'' and ``31'' entries can be removed by choosing 
$p_1=p_2=0$ and $p_3 =\pi$, the resultant matrix can be written in a more familiar forms
\begin{eqnarray}
m_\nu &=& 
\left (\begin{array}{ccc}
A& C& C^*\\
C& D^*&\tilde B\\
C^*&\tilde B& D
\end{array}
\right )\;,\label{real}
\end{eqnarray}
where $\tilde B = - B$.

The simplicity of the above mass matrix may serve as a good starting point to understand the possible underlying theory.  If this has something to do with reality, one should not stay at the pure phenomenological level for analysis, but go further to study whether there are theoretical models which can obtained such a neutrino mass matrix in some consistently way. Several attempts for model buildings have been made\cite{ma-a4,grimus}.  It has been shown in ref. \cite{gls} by Grimus and Lavoura that the above form of mass matrix is symmetric under a transformation of $e\to e$, $\mu-\tau$ exchange with a CP conjugation. We will refer this as the Grimus-Lavoura symmetry (GLS).
In this work we start with a simple model proposed earlier based on $A_4$ symmetry\cite{hev} to realize the tri-bimaximal neutrino mixing, and then modify it to allow a non-zero $\theta_{13}$ to find the conditions for having the GLS limit for neutrino mass matrix with $\delta = -\pi/2$ and $\theta_{23} = \pi/4$ and how modifications may occur by explicit model studies. This model has an added bonus that\cite{hezee,hev} $s_{12}c_{13}= V_{e2} = 1/\sqrt{3}$. There is also an interesting feature in this model that CP violation can be solely from complexity of relevant Clebsh-Gordan (C-G) coefficients in the GLS limit. We will refer this property as intrinsic CP violation. 

In this model $A_4$ is serving as a family symmetry\cite{hev}. The Higgs sector is enlarged to have three Higgs
fields, $\Phi = (\Phi_1,\;\Phi_2,\;\Phi_3)$ (SM doublet), $\phi$ (SM doublet) and
$\chi= (\chi_1,\;\chi_2,\;\chi_3)$ (SM singlet). Under the $A_4$, $\Phi$ and
$\chi $ both transform as 3, and $\phi$ as 1. Three right-handed SM singlet neutrinos $\nu_R=(\nu_R^1,\;\nu_R^2,\;\nu_R^3)$ are introduced allowing seesaw mechanism to be in effective.
The standard
left-handed leptons $l_L = (l_L^1,\;l_L^2,\;l_L^3)$, and standard right-handed charged leptons $(l^1_R,
l^2_R, l^3_R)$, and $\nu_R$ transform as a 3 ,
$(1,1'',1')$ and 3, respectively. We refer the readers for more
details on $A_4$ group properties to Refs.\cite{ma-a4,hev,zee}.
The Lagrangian responsible for the lepton mass matrix is
\begin{eqnarray}
L &=& \lambda_e (\bar l_L \tilde \Phi)_1 l^1_R + \lambda_\mu (\bar l_L
\tilde \Phi)_{1'} l^2_R + \lambda_\tau (\bar l_L \tilde \Phi)_{1''}
l^3_R + H.C.\nonumber\\
&+& \lambda_\nu (\bar l_L \nu_R)_1 \phi + m (\bar \nu_R \nu^C_R)_1 +
\lambda_\chi (\bar \nu_R \nu^C_R)_3 \chi,
\end{eqnarray}
where
\begin{eqnarray}
&(\bar l_L \tilde \Phi)_1 l^1_R&= (\bar l^1_L \tilde\Phi_1 + \bar l^2_L \Phi_2 +\bar l_L^3\tilde \Phi_3)l_R^1\;,\nonumber\\
&(\bar l_L \tilde \Phi)_{1'} l^1_R&= (\bar l^1_L \tilde\Phi_1 + \omega \bar l^2_L \Phi_2 +\omega^2\bar l_L^3\tilde \Phi_3)l_R^2\;,\\
&(\bar l_L \tilde \Phi)_{1''} l^1_R&= (\bar l^1_L \tilde\Phi_1 + \omega^2\bar l^2_L \Phi_2 +\omega\bar l_L^3\tilde \Phi_3)l_R^3\;.\nonumber
\end{eqnarray}
Here $\omega = exp(i 2\pi/3)$ and $\omega^2 = exp(i4\pi/3)$ are the C-G coefficients of the $A_4$ group products.

If the vev structure is of the form $<\Phi_{1,2,3}> = v_\Phi$,
$<\chi_{1,3}>=0$, $<\chi_2> = v_\chi$, and $<\phi> = v_\phi$, one
would obtain the charged lepton mass term as
\begin{eqnarray}
&&\left (\begin{array}{lll} \bar l^1_L& \bar l^2_L& \bar l^3_L
\end{array} \right )U_l \left ( \begin{array}{lll}
\sqrt{3}\lambda_e v_\Phi&0&0\\0&\sqrt{3}\lambda_\mu v_\Phi&0\\0&0&\sqrt{3}\lambda_\tau
v_\Phi\end{array}\right )\left (
\begin{array}{r}l^1_R\\l^2_R\\ l^3_R\end{array}\right )\;, \;\;U_l = {1\over \sqrt{3}}\left ( \begin{array}{lll}
1&1&1\\
1&\omega&\omega^2\\
1&\omega^2&\omega
\end{array}\right )\;.
\end{eqnarray}
From the above, we can identify the charged lepton mass to be
$m_i =\sqrt{3}\lambda_i v_\Phi$.
The neutrino mass matrix has the seesaw form with
\begin{eqnarray}
M &=& \left ( \begin{array}{ll} 0&M_D \\M_D^T
&M_R\end{array}\right ),\;\; M_R =\left (
\begin{array}{lll}m&0&m_\chi\\0&m&0\\m_\chi&0&m\end{array}\right
),
\end{eqnarray}
where $M_D = Diag(1,1,1)\lambda_\nu v_\phi$, and $m_\chi =
\lambda_\chi v_\chi$. From this one obtains the light neutrino
mass matrix $M_\nu$ of the form given by
\begin{eqnarray}
M_\nu = - M_DM_R^{-1}M_D = \left ( \begin{array}{lll}
w&0&x\\
0&y&0\\
x&0&z
\end{array}\right )\;,
\end{eqnarray}
where $w =z= -(\lambda_\nu v_\phi)^2m/(m^2-m_\chi^2)$, $x=(\lambda_\nu v_\phi)^2m_\chi/(m^2-m_\chi^2)$ and
$y = -(\lambda_\nu v_\phi)^2/m$.

The above model leads to the tri-bimaximal mixing which predicts $\theta_{13}=0$. It had been the focus of $A_4$ symmetry studies for a few years\cite{alterali-hebabu,hev, he1}. But it is now ruled out because a non-zero $\theta_{13}$ has been measured. In this scheme, in order to obtain the tri-bimaximal mixing, 
the neutrino mass matrix with ``11'' and ``33'' entries to be equal is crucial. It has been pointed out\cite{hev} that a more natural form of vev structure will lead to the ``33'' entry in the neutrino mass matrix to be deviate from the ``11'' entry which leads to a non-zero $\theta_{13}$. To achieve this, for our purpose
here, we will introduce two scalars $S_{1'}$ and $S_{1''}$ which are SM singlet but transform as $1'$ and $1''$ under $A_4$. This results in two new terms for $M_R$ in the Lagrangian
\begin{eqnarray}
Y_{S'} (\bar \nu_R \nu_R^C)_{1''}S_{1'} + Y_{S''} (\bar \nu_R \nu_R^C)_{1'}S_{1''} + H.C.
\end{eqnarray}
After $S_{1',1''}$ develops a non-zero vev, $v_{S',S''}$, we have
\begin{eqnarray}
M_R =\left (
\begin{array}{lll}m_1 &0&m_\chi\\0&m_2 &0\\m_\chi&0&m_3 \end{array}\right
),
\end{eqnarray}
where $m_1= m+ Y_{S'}v_{S'} + Y_{S''} v_{S''}$, $ m_2= m+ \omega^2Y_{S'}v_{S'} + \omega Y_{S''} v_{S''}$ and $m_3 =m+ \omega Y_{S'}v_{S'} + \omega^2 Y_{S''} v_{S''}$. 
The resulting light neutrino mass matrix $M_\nu$ no longer has $w=z$, but has 
\begin{eqnarray}
w= -\lambda^2_\nu v^2_\phi m_3/(m_1 m_2 - m^2_\chi)\;,\;\;\;\;
z=- \lambda^2_\nu m_1/(m_1m_3 - m^2_\chi)\;,
\end{eqnarray}
and $x$ and $y$ are changed to
\begin{eqnarray} 
x =\lambda_\nu^2v_\phi^2m_\chi /(m_1 m_3 - m^2_\chi)\;,\;\;\;\; y=-\lambda_\nu^2v_\phi^2 /m_2\;.
\end{eqnarray}
In the basis where the charged lepton mass matrix is diagonalized, the neutrino mass matrix becomes
\begin{eqnarray}
m_\nu = U_l^\dagger M_\nu U^*_l = {1\over 3}\left (\begin{array}{ccc}
w+2x+y+z&w-\omega^2 x +\omega^2 y +\omega z& w-\omega x +\omega y +\omega^2 z\\
w-\omega^2 x +\omega^2 y +\omega z&w+2\omega x +\omega y +\omega^2 z&w-x+y+z\\
w-\omega x +\omega y +\omega^2 z&w-x+y+z&w+2\omega^2 x +\omega^2 y +\omega z
\end{array}
\right )\;.
\label{model-matrix}
\end{eqnarray}

Inserting $\omega = exp(i2\pi/3)$ in the above, $m_\nu$ can be transformed into the form in eq.(\ref{matrix}) by redefine right-handed charged leptons. The parameters in the set $P_{A4}: \{w, x, y, z \}$ are in general complex which will not always have $\delta = -\pi/2$ and $\theta_{23} = \pi/4$. One needs to work in the GLS limit which can be realized if the parameters in the set $P_{A4}$ are all real. 
In this case the complexity of the mass matrix is purely due to the $A_4$ group theoretical C-G coefficients $\omega$ and $\omega^2$. This is a case where CP violation is caused by C-G coefficients providing a concrete example of intrinsic CP violation. 

Before we analysis the general features of the neutrino mass matrix with complex parameters in the set $P_{A_4}$, we would like to analysis the constraints on the model parameters to have the GLS limit, that is, to have $w, x, y, z$ to be real. The complexity of the parameters can appear in the Yukawa couplings, in the vevs, and also in places where $\omega^i$ appear in $m_i$. 
To make the Yukawa couplings and scalar vevs real, one can require the model Lagrangian to satisfy a generalized CP symmetry under which
\begin{eqnarray}
&&(l^1_L \;,\;\;l^2_L\;,\;l^3_L) \to ( (l^1_L)^{CP}\;,\;\; (l^3_L)^{CP}\;,\;\;(l^2_L)^{CP})\;,\;\;
(\nu^1_R \;,\;\;\nu^2_R\;,\;\nu^3_R) \to ( (\nu^1_R)^{CP}\;,\;\; (\nu^3_R)^{CP}\;,\;\;(\nu^2_R)^{CP})\;,
\nonumber\\
&&(\Phi_1, \Phi_2, \Phi_3) \to (\Phi^\dagger_1, \Phi^\dagger_3, \Phi^\dagger_2)\;,\;\;(\chi_1\;,\;\chi_2\;,\;\;\chi_3) \to (\chi_1^\dagger\;, \;\;\chi_3^{\dagger}\;,\;\; \chi_2^\dagger)\;,\;\;(S_{1'}\;,\;\;S_{1''}) \to (S^\dagger_{1'}\;,\;\;S^\dagger_{1''})\;,
\end{eqnarray}
and all other fields transform the same as those under the usual CP symmetry. Here the superscript $CP$ in the above indicates that the fields are the usual $CP$ transformed fields.

The above transformation properties will transform relevant terms into their complex conjugate ones. Requiring the Lagrangian to be invariant under the above transformation dictates the Yukawa couplings to be real. The same requirement will dictates the scalar potential to forbid spontaneous CP violation and vevs to be real. One, however, notices that the parameters
$m_{2,3}$ are in general complex even if the Yukawa couplings and the vevs of the scalar fields are made real because of the appearance of $\omega^i$. To make them real to reach GLS limit, it is therefore required that
\begin{eqnarray}
Im(\omega^2 Y_{S'} v_{S'}+\omega Y_{S''} v_{S''})=Im(\omega Y_{S'} v_{S'}+\omega^2 Y_{S''} v_{S''}) = 0\;.
\end{eqnarray}
The above can be achieved by the absent of the scalar fields $S^{',''}$ in the theory or set $Y_{S'} v_{S'}= Y_{S''}v_{S''}$. 
If the vev structure of $\chi$ is fixed as given previously, absence of $S^{',''}$ will not have a phenomenologically acceptable mass matrix. Therefore, we will take the later possibility as example of GLS limit case to show some detailed features. In this case $M_\nu$ can be diagonalized by $V_\nu$ as the following
\begin{eqnarray}
M_\nu = V_\nu \hat m_\nu V_\nu^T\;, \;\;V_\nu = \left (\begin{array}{ccc}
c&0&-s\\
0&1&0\\
s&0&c
\end{array}
\right )\;,
\end{eqnarray}
where $s = \sin\theta$ and $c = \cos\theta$. 
One obtains the mixing matrix to be
\begin{eqnarray}
V_{PMNS} = U_l^\dagger V_\nu &=& {1\over \sqrt{3}}\left(
\begin{array}{ccc}
c + s & 1 &
c-s \\
c+\omega s& \omega^2 &
\omega c - s \\
c+\omega^2 s& \omega & \omega^2 c -s
\end{array}
\right)\;,
\label{real-mixing}
\end{eqnarray}
Normalizing the above mixing matrix to the standard parametrization in eq.(\ref{pmns}), one obtains
\begin{eqnarray}
s_{12} ={1\over \sqrt{2(1+cs)}}\;,\;\;s_{23} = {1\over \sqrt{2}}\;,\;\;s_{13} = {(1-2cs)^{1/2}\over \sqrt{3}}\;.
\end{eqnarray}
Here we have normalized $c_{ij}$ and $s_{ij}$ to be all positive. The neutrino eigen-masses are all real, but in general they can take positive or negative values depending on the values of $w$, $x$, $y$ and $z$. Note that the absolute values of elements in the second column of $V_{PMNS}$ are all $1/\sqrt{3}$.

We now find the conditions for predicting $\delta = -\pi/2$ and $\delta = +\pi/2$.
An easy way of doing this is to study the Jarlskog invariant quantity\cite{jarlskog} $J = Im(V_{e1}V^*_{e2}V_{\mu 1}^*V_{\mu 2})$.
Eqs.(\ref{pmns}) and (\ref{real-mixing}) give
\begin{eqnarray}
J = c_{13}^2s_{12}c_{12}s_{23}c_{23}s_{13} sin\delta = -{1\over 6\sqrt{3}}(c^2-s^2)\;,
\end{eqnarray}
which leads to 
\begin{eqnarray}
\delta ={\pi\over 2} \times \left \{ \begin{array} {l}
-1\;,\;\; \mbox{if}\;\;c^2>s^2\;,\\
+1\;,\;\;\mbox{if}\;\;s^2>c^2\;.
\end{array} \right .
\end{eqnarray}

Note that $J$ is not zero implying CP violation which is caused by the complexity of C-G coefficients.
Eq. (\ref{real-mixing}) can be transformed into the standard parameterization by multiplying the $V_{PMNS}$ on the right and left
by diagonal matrices $P_r = diag(1, 1, i)$ and $P_l = diag(1, (\omega^2 c-s)/|\omega^2 c -s|, (\omega c -s)/|\omega c -s|)$, respectively. $P_l$ does not have physical effect because it can be absorbed by redefinition of right-handed charged leptons. The physical effects of $P_r$ is to change the sign of $m_3$. 

Let us now compare experimental data with the model predictions for the mixing angles and CP violating phase. There are several global fits of neutrino data\cite{valle,fogli-schwetz}. The latest fit gives the central values, 1$\sigma$ errors and the 2$\sigma$ ranges as the following\cite{valle}
\begin{eqnarray}
\begin{array}{ccccc}
&\delta/\pi&s_{12}^2&s_{13}^2/10^{-2}&s_{23}^2\\
NH&1.41^{+0.55}_{-0.44}&0.323\pm 0.016 &2.26\pm 0.12&0.567^{+0.032}_{-0.124}\\
2\sigma\;\mbox{region}&0.0\sim 2.0&0.292 \sim 0.357&2.02 \sim 2.50&0.414\sim 0.623\\
IH&1.48\pm 0.31&0.323\pm0.016&2.29\pm0.12&0.573^{+0.025}_{-0.039}\\
2\sigma\;\mbox{region}&0.00\sim 0.09 \& 0.86 \sim 2.0&0.292\sim 0.357&2.05\sim 2.52&0.435\sim 0.621
\end{array}
\end{eqnarray}
Here $NH$ and $IH$ indicate neutrino mass hierarchy patterns of normal hierarchy and inverted hierarchy, respectively.
In the model above, adjusting the values, $w$, $x$, $y$ and $z$, both NH and IH mass patterns can be obtained. There is strong hint that the Dirac phase should be close to $3\pi/2$(or equivalently $-\pi/2$). Therefore one should take the parameter space so that $c^2>s^2$. The value $-\pi/2$ predicted in the model is in agreement with IH within 1$\sigma$ range. Although for NH case $\delta$ is outside of 1$\sigma$ range, there no problem with 2$\sigma$ range. For $s_{23}$, the model predicts $s^2_{23} = 0.5$. This value is outside of 1$\sigma$ range for both the NH and IH cases. However, they are, again, in agreement with data within 2$\sigma$.

In the model $s_{13} = (1-2cs)^{1/2}/\sqrt{3}$ is not predicted. But one can use information from $s_{13}$ to fix 
$cs = 0.497\pm 0.018$ to predict $s_{12}^2= 0.334\pm0.004$ for both NH and IH cases. This is in agreement with data within 1$\sigma$. Note that $V_{e2}^2 = (s_{12}c_{13})^2 = 1/3$. It agrees with data within 1$\sigma$.
It is remarkable that neutrino mixing matrix in this model with just one free parameter can be in reasonable agreement with data. This may be a hint that it is the form for mixing matrix, at least as the lowest order approximation, that a underlying theory is producing.

If $w$, $x$, $y$ and $z$ are allowed to be complex, the GLS is explicitly broken, there are modifications to the mixing angles. There is additional source for CP violation other than the intrinsic one from complexity of C-G coefficient, and also the mixing angles will be modified. The eigen-masses will contain Majorana phases. Detailed analysis of how to diagonalize the mass matrix has been discussed in Ref.\cite{hezee}. In general this model does not always predicts $\delta = \pm\pi/2$ and $\theta_{23} = \pi/4$. The mixing matrix can be, in general, written as
\begin{eqnarray}
V_{PMNS} = U_l^\dagger V_\rho V_\nu = {1\over \sqrt{3}}\left(
\begin{array}{ccc}
c + se^{i\rho} & 1 &
ce^{i\rho}-s \\
c+\omega se^{i\rho}& \omega^2 &
\omega ce^{i\rho} - s \\
c+\omega^2 se^{i\rho}& \omega & \omega^2 ce^{i\rho} -s
\end{array}
\right)\;,
\end{eqnarray}
where $V_\rho$ is a diagonal matrix $diag(1,1,e^{i\rho})$ with $\tan\rho = Im (xw^*+x^*z)/Re(xw^*+x^*z)$.
It is interesting that the phase $\rho$ does not show up in $J$ which is still 
$-(c^2-s^2)/6\sqrt{3}$. This implies that CP violation related to neutrino oscillation is still purely due to intrinsic CP violation. The mixing angles and the Dirac phase $\delta$ are all modified with
\begin{eqnarray}
s_{12} = {1\over \sqrt{2}(1+cs\cos\rho)^{1/2}}\;,\;s_{23} = {(1+cs \cos\rho + \sqrt{3} cs \sin\rho)^{1/2}\over \sqrt{2}(1+cs \cos\rho)^{1\over 2}}\;,\;s_{13} = {(1-2cs \cos \rho)^{1/2}\over \sqrt{3}}\;,
\end{eqnarray}
and 
\begin{eqnarray}
\sin\delta = (1+{4 c^2s^2 \sin^2\rho\over (c^2-s^2)^2)})^{-1/2}(1- {3 c^2s^2\sin^2\rho\over (1+cs \cos\rho)^2})^{-1/2} \times 
\left \{ \begin{array} {l}
-1\;,\;\; \mbox{if}\;\;c^2>s^2\;,\\
+1\;,\;\;\mbox{if}\;\;s^2>c^2\;.
\end{array} \right .
\end{eqnarray}

In this case, the new parameter $\rho$ can be used to improve agreement of the model with data.
In both NH and IH cases, $\delta$ and $s_{23}$ can be brought into agreement with data at 1$\sigma$ level. 
To see how this can be done, as an example, we take the largest value of $cs$ so that $s_{13}$ takes its lower 1$\sigma$ allowed value, and then varying $\cos\rho$ to obtain the upper 1$\sigma$ allowed value. This fixes $cs$ and $\cos\rho$ to be 
0.468 and 0.992, respectively. With these values, $s_{23}$ and $\delta$ are determined to:
$0.534$ and $1.426 \pi$, respectively. These values are in agreement with data at 1$\sigma$ level. 
When more precise experimental data become available, the model with complex model parameters can be distinguished from that with the parameters are all real and other models.

In summary we have shown that neutrino mass matrix reconstructed with $\delta = -\pi/2$ and $\theta_{23} = \pi/4$ has several interesting properties. We find that a theoretical model based on the $A_4$ symmetry naturally realize the GLS limit and predicts such a neutrino mixing pattern together with the prediction $|V_{e2}|=1/\sqrt{3}$. In this model, CP violation can be solely come from the complex group theoretical C-G coefficients if the neutrino Majorana phases are zero or $\pi$. This model fits experimental data very well and can be taken as the lowest order neutrino mass matrix for future theoretical model buildings. If there are additional source of CP violation other than those intrinsically existed in the C-G coefficients, the CP violating phase $\delta$ and the mixing angle $\theta_{23}$ can be away from $-\pi/2$ and $\pi/4$. The models discussed can fit data within 1$\sigma$. Future improved experimental data will be able to further test the model and provide more hints for the underlying theory of neutrino mixing. 

\begin{acknowledgments}
The work was supported in part by MOE Academic Excellent Program (Grant No: 102R891505) and MOST of ROC, and in part by NSFC(Grant No:11175115) and Shanghai Science and Technology Commission (Grant No: 11DZ2260700) of PRC.
\end{acknowledgments}

\end{document}